\documentstyle[12pt,titlepage,epsf]{article}
\title{IMPROVED CONSTRAINTS ON $\bar B^{0} \rightarrow \pi^{+} \ell^{-} \bar 
\nu_{\ell}$ FORM FACTORS}
\author{Damir Becirevic}
\date{February, 1996} 
\def\bpi{B \rightarrow \pi\ell \bar\nu_{\ell} }
\begin{document} 
\begin{titlepage}
\begin{flushright}
LPTHE - Orsay 96/14 
hep-ph/9603298
\end{flushright}
\begin{center}
\vspace{0.75in}
  {\bf {\LARGE Improved Constraints on $\bpi$  Form Factors}} \\
  \vspace{0.75in}
Damir Becirevic
\vskip 0.5cm
{\small \it Laboratoire de Physique Th\'eorique et Hautes Energies\\
Universit\'e de Paris XI, B\^atiment 211, 91405 Orsay Cedex, France}\\
\vskip 0.8in
  ABSTRACT
 \vspace{0.8cm}
  \end{center}
\begin{quotation}
  \noindent

The behavior of the $\bpi$ form factors in the entire physical range is 
examined in a model independent way. Unitarity bounds are further 
constrained by the lattice results (APE). From this analysis, we obtain 
$f^{+}(0) = 0.38 \pm .03$ if the $B^{*}$-pole dominance behavior is assumed. 
However, to get the information on the behavior of the form factor $f^{+}$, 
QSR results are included. We see the deviations from the pole dominance due 
to the contributions of the higher singularities which are to be treated 
with more precise lattice data.  
\end{quotation}
\end{titlepage}
\newpage 
\section{Introduction}
In this paper we would like to examine the $\bpi$ form-factors in the entire 
physical region in a model independent way. This is, of course, very 
important for the forecoming B-factories' experiments which will provide us 
a way to extract the precise value of $V_{ub}$. This CKM- matrix element 
plays the crucial role in our understanding of the mechanism of $CP$ 
violation. $V_{ub}$ is mainly determined from the end-point of the lepton 
spectrum in semileptonic $B$-decays. Unfortunately, the theory of the 
end-point region of the lepton spectrum in inclusive $B \rightarrow X_{u} 
\ell \bar \nu_{\ell}$ decays is very complicated and suffers from large 
uncertainties. Hence, we explore exclusive decay modes which are also easier 
for experimentalists. The natural candidate is $\bpi$.
However, the determination of $V_{ub}$ depends on the knowledge of the 
physics at large distances, i.e. non-perturbative $QCD$.
This problem has proved to be notoriously difficult. We cannot rely on the 
heavy quark symmetry (HQS) to reduce number of the form factors or to fix 
their normalisationas at $q^{2}_{max}$. In heavy - heavy transition, heavy 
quark symmetry helps to calculate the normalisation at the zero-recoil point 
with small and controlled theoretical errors. This is the crucial point for 
the model independent determination of $|V_{cb}|$. For the case of the light 
resulting particle, we have no such a guidance. The knowledge of the 
form-factor at several points does not mean the knowledge of its functional 
behavior. Existing experimental values ($CLEO$) \cite{cleo} for the 
branching ratio are model dependent ($WBS$ \cite{wsb} and Isgur-Wise 
\cite{isgw} models):
\begin{equation}
 Br(\bar B^{0} \rightarrow \pi^{+} \ell^{-} \bar \nu_{\ell}) = \left\{  
\begin{array}{c c}
(1.63 \pm 0.46 \pm 0.34) 10^{-4} & WSB \\
(1.34 \pm 0.35 \pm 0.28) 10^{-4} & ISGW 
\end{array} \right.
\end{equation}

The problems with these models are related to the fact that they are always 
not relativistic in some aspects and they do not provide real predictions 
for the form factors. Small recoil behavior of the form factor was also 
calculated by  combined  HQS and the chiral perturbation theory \cite{ch}. 
This approach is model independent, but it is valid only in the soft-pion 
limit. However, there are two methods that are rooted in the QCD from the 
first principles: the lattice QCD ($LQCD$) and the $QCD$ sum rules ($QSR$). 
While the lattice can be employed to explore the region close to the 
zero-recoil point (i.e. near $q^{2}_{max}$), the QSR give us the value of 
the form-factors for small $q^{2}$. Unfortunately, the regions where these 
two methods apply do not 
overlap so that the intermediate region stays uncovered (at least not with a 
considerable precision). So far, in the lattice approach the values of 
form-factors were calculated at several points (several $q^{2}$), and then 
extrapolated according to an ansatz of functional dependence on $q^{2}$. 
With faster computers and new simulations, the extrapolations will be far 
more constrained. For the moment, we would like to obtain the bounds on the 
form factors, and in a consistent way treat the existing results in order to 
constrain these bounds. In a beautiful series of papers, Boyd, Grinstein and 
Lebed \cite{boy,gri,leb} applied an old method \cite{oku} to the heavy quark 
systems. The idea is to use crossing symmetry and dispersion relations in 
order to relate form factors with QCD perturbative calculations in an 
unphysical kinematic region. To constrain the bounds obtained in this way, 
we shall use some well estimated LQCD-results (APE), as well as some 
QSR-results.
The paper is organised as follows: In Sec.2 we outline the basic theoretical 
formalism and obtain the bounds. In Sec.3 we strengthen these bounds by using
existing results and  make an short analysis of the $q^2$ behavior of the 
form-factors.

\section{General Formalism}
For the sake of completeness, in this section we recall the main features of 
the method ( see references \cite{oku,gri,raf,leb,boy})
The current matrix element governing the $\bpi$ semileptonic decay is 
parametrized as

\begin{eqnarray}
<\pi (p')|V^\mu (0)|B(p)> = (p+p'-q \frac{m_{B}^2 -m_{\pi}^2}{q^{2}})^{
\mu}f^{+}(q^{2}) + q^{\mu}\frac{m_{B}^2 -m_{\pi}^2}{q^{2}} f^{0}(q^{2}) 
\end{eqnarray}
where $V^\mu=\bar u\gamma^{\mu}b$, and the form factors $f^{+,0}(q^{2})$ are 
functions of the momentum transfer $t \equiv q^{2} = (p - p')^{2}$. They 
satisfy the kinematical constraint: $f^{+}(0)=f^{0}(0)$. Here, we assume 
that leptons are light so that the physical region of t accessible from this 
decay is $0 \leq t \leq (M-m)^{2}$.
The expression for the decay rate is then:
\begin{eqnarray}
\frac{d\Gamma}{dq^{2}}(\bar B^{0} \rightarrow \pi^{+} \ell^{-} \bar \nu_{
\ell}) = \frac{G^{2}|V_{ub}|^{2}}{192\pi^{3}m_{B}^{3}}
\lambda^{3/2}(q^{2})|f^{+}(q^{2})|^{2}
\end{eqnarray}
where $\lambda (t) = (t + m_{B}^{2} - m_{\pi}^{2})^{2} - 4 m_{B}^{2} m_{
\pi}^{2}$ is the usual triangular function.
If we want to extract the precise value of $V_{ud}$, it is obvious how 
important is the knowledge of the functional behavior of $f^{+}(t)$.
To derive the bounds we consider the two point function:
\begin{eqnarray}
\Pi ^{\mu \nu} \equiv i\int d^{4}x e^{iqx} <0|T(V^\mu (x)V^{\nu
\dagger}(0))|0> = 
(q^{\mu}q^{\nu} - q^{2}g^{\mu \nu})\Pi_{T}(q^{2}) + g^{\mu \nu}\Pi_{L}(q^{2})
\end{eqnarray}
 
In the case of $QCD$, the structure functions $\Pi_{T,L}(q^{2})$ satisfy the 
once 
subtracted dispersion relations:
\begin{equation}
\chi_{T,L}(Q^{2}) = \frac{\partial \Pi_{T,L}(q^{2})}{\partial q^{2}} 
\mid_{q^{2}=-Q^{2}} = \frac{1}{\pi} \int_{0}^{\infty} \frac{Im 
\Pi_{T,L}(t)}{(t + Q^{2})^{2}}dt
\end{equation}
The functions $\chi_{T,L}(Q^{2})$ can be reliably calculated in perturbative 
QCD as long as we stay in the region far from resonances. So, it 
suffices to calculate them at $Q^2=0$ where $(m_{b} + m_{u})\Lambda_{QCD} << 
(m_{b} + m_{u})^{2} + Q^{2}$ is satisfied. To one loop they read:
\begin{equation}
\chi_{T}(0) = \frac{1}{8\pi^{2}(m_{b}^{2}-m_{u}^{2})^{5}}\left[ 
(m_{b}^{4}-m_{u}^{4})(m_{b}^{4}+m_{u}^{4}-8m_{b}^{2}m_{u}^{2}) - 
12m_{b}^{4}m_{u}^{4} \log{\frac{m_{b}^{2}}{m_{u}^{2}}} \right]
\end{equation}

\begin{equation}
\chi_{L}(0) = \frac{
(m_{b}^{2}-m_{u}^{2})(m_{b}^{2}+m_{u}^{2} + m_{b}m_{u})(m_{b}^{2}+m_{u}^{2} 
- 4m_{b}m_{u}) -
6m_{b}^{3}m_{u}^{3} \log{ \frac{m_{b}^{2}}{m_{u}^{2}}}} {8\pi^{2}(m_{b}^{2} 
+m_{u}^{2})^{3}}
\end{equation}
For a massless $u$-quark, it gives: $\chi_{T}(0)=1.27\cdot 
10^{-2}/m_{b}^{2}$ 
and $\chi_{L}(0)=1.90\cdot 10^{-4}$ (from now on, we shall simply note $
\chi_{T,L}$, instead of $\chi_{T,L}(0)$). Also, the $O(\alpha_{s})$ 
corrections can 
be included \cite {snar}and they enhance the above expressions for less than 
$20\%$.

The absorptive parts of the spectral functions $Im \Pi_{L,T}(q^{2})$ can be 
obtained by inserting the on-shell states between the two currents on the 
l.h.s. of Eq.4. taking the $|\bar B\pi>$ as the lowest state 
contributing to the absorptive amplitude.  Crossing symmetry states that the 
matrix element is described by the same form factors, but real for $t_{+}
\leq t\leq \infty$. After integration over the phase-space, the longitudinal 
part becomes: 
\begin{equation}
Im \Pi_{L}(q^{2}) \geq \frac{[(t-t_{+})(t-t_{-})]^{1/2}}{16 \pi t^{2}} 
t_{+}t_{-}|f^{0}(t)|^{2}\theta (t - t_{+}) \\ 
\end{equation}
Below the onset of the $B\pi$-continuum there is only one resonance - 
$B^{*}(1^{-})$, which according to its quantum numbers contributes to $Im 
\Pi_{T}$, which reads:
\begin{eqnarray}
Im \Pi_{T}(q^{2})\geq \pi f_{B^{*}}^{2} \delta(t - m_{B^{*}}^{2})
+ \frac{[(t-t_{+})(t-t_{-})]^{3/2}}{48 \pi t^{3}}|f^{+}(t)|^{2}\theta (t - 
t_{+})  
\end{eqnarray}
where $t_{\pm}=m_{B}\pm m_{\pi}$. Thus, replacing the absorptive parts in 
dispersion relations, we obtain the following inequalities:
\begin{eqnarray}
\chi_{L} &\geq& \int_{t_{+}}^{\infty} \varphi_{L}(t)
|f^{0}(t)|^{2}dt , \cr
\chi_{T} &\geq& \frac{f_{B^{*}}^{2}}{m_{B^{*}}^{4}} + \int_{t_{+}}^{\infty} 
\varphi_{T}(t)|f^{+}(t)|^{2}dt ,
\end{eqnarray} 
where we put $t^{-2} Im \Pi_{i}(t)\equiv \varphi_{i}(t) |f_{i}(t)|^{2} 
\theta 
(t - t_{+})$. After integration, the pole-contribution becomes very small 
($f_{B^{*}}^{2}/m_{B^{*}}^{4}$), and can be safely neglected. We absorb this 
contribution intoto $\chi_{T}$, anyway.
To get informations about the form factors in the physical region, we 
map the complex t-plane onto the unit disc $|z|\leq 1$:
 
\begin{eqnarray}
\frac{1+z}{1-z} = \sqrt{\frac{(m_{B}+m_{\pi})^{2}-t}{4m_{B}m_{\pi}}}
\end{eqnarray}
By this transformation, the region $t_{-}\leq t\leq t_{+}$ is mapped into 
the segment of the real axis $-1<z\leq 0$, while the $0 \leq t\leq t_{-}$ is 
mapped into $0\leq z<1$. Two branches of the root (or two sides of the cut) 
are mapped to upper and lower semicircles of $|z|=1$. Physically, the 
relevant kinematic region for the process $vacuum \to \bar B\pi$ now lies on 
the unit circle, while the region for the semileptonic $\bpi$ decay lies 
inside the unit circle, on the real axis.  Generically rewritten in the 
$z$-plane, inequalities (10) are:
\begin{eqnarray}
\frac{1}{2\pi i}\int_{{\cal C}:|z|=1}\frac{dz}{z}|\phi_{i}(z)f_{i}(z)|^{2} 
\leq \chi_{i}
\end{eqnarray}
The functions $\phi_{i}(z)$ are solutions of the Dirichlet's boundary 
problem \cite{raf} of finding an analytic function on the unit disc. Their 
values are known on the circle: $|\phi_{i}(e^{i\theta})|^{2}=\varphi_{i}(e^{i
\theta})$, where $\varphi_{i}$ are the transformed functions from the 
integrals (9) . The solution is:
\begin{eqnarray}
\log{|\phi (z)|^{2}} = \frac{1}{2\pi }\int_{0}^{2\pi}d\theta  \frac{e^{i
\theta}+z}{e^{i\theta}-z} \log{\varphi(e^{i\theta})}
\end{eqnarray}
Explicitly, our functions are:
\begin{eqnarray}
\phi_{T} (z) = \frac{1}{\sqrt{6\pi m_{B}m_{\pi}}}\frac{(1+z)^2}{(1-z)^{9/2}} 
\left( \frac{m_{B}+m_{\pi}}{2\sqrt{m_{B}m_{\pi}}} + \frac{1+z}{1-z}\right) 
^{-5}
\end{eqnarray}
\begin{eqnarray}
\phi_{L} (z) = \frac{m_{B}^{2}-m_{\pi}^{2}}{8m_{B}m_{\pi}\sqrt{2\pi m_{B}m_{
\pi}}} \frac{1+z}{(1-z)^{5/2}} \left( \frac{m_{B}+m_{\pi}}{2\sqrt{m_{B}m_{
\pi}}} + \frac{1+z}{1-z}\right) ^{-4}
\end{eqnarray}
Still, the $B^{*}$-pole ($z_{pole}=-0.2519$) is below threshold and cannot 
be 
ignored when we consider $f^{+}$. We do not know the size of the residue, 
but the $m_{pole}$ is known. So, to remove the pole, we multiply $f^{+}(z)$ 
by the Blaschke factor:
\begin{eqnarray}
P_{*}(z) = \frac{z-z_{pole}}{1-z  z_{pole}^{*}}
\end{eqnarray}
which is unimodular on the unit circle and principally does not spoil our 
analysis although it 
will slightly weaken our  bounds. Now, the product $
\phi_{i}(z)f_{i}(z)P_{*}(z)$ is analytic on the unit disc and obeys (12). It 
should be noted 
that in our case, there are no branch points below the threshold.
The last step in deriving the bounds is to construct the inner product:  
\begin{equation}
(g_{1},g_{2}) = \int_{\cal C} \frac{dz}{2\pi iz}  g_{1}^{*}(z) g_{2}(z).
\end{equation}
Let us choose $g_{1}(z)=\phi_{i}(z)f_{i}(z)$ and $g_{2}(z)=(1 - 
zz_{2}^{*})^{-1}$. From the positivity of the inner product, determinant of 
the $(g_{i},g_{j})$ matrix is positive, so that from these two functions we 
have:
\begin{equation}
\left| \begin{array}{cc}
\chi_{i} & f_{i}^{*}(z_{2})\phi_{i}^{*} (z_{2}) \\
f_{i}(z_{2})\phi_{i}(z_{2}) & \frac{1}{1 - |z_{2}|^{2}}
\end{array} \right|  \geq 0  ,  
\end{equation}
$\forall z_{2}\in {\mathrm Int} \cal C$ \\
or explicitly
\begin{equation}
|f_{i}(z)|^{2} \leq \frac{1}{|\phi_{i}(z)|^{2}}\frac{\chi_{i}}{1 - |z|^{2}}
\end{equation}
For the case of $f^{0}(z)$ these bounds are depicted on the Fig.1. Coming 
from general principles, the bounds obtained in this way cannot be strong.
$$ \epsfbox{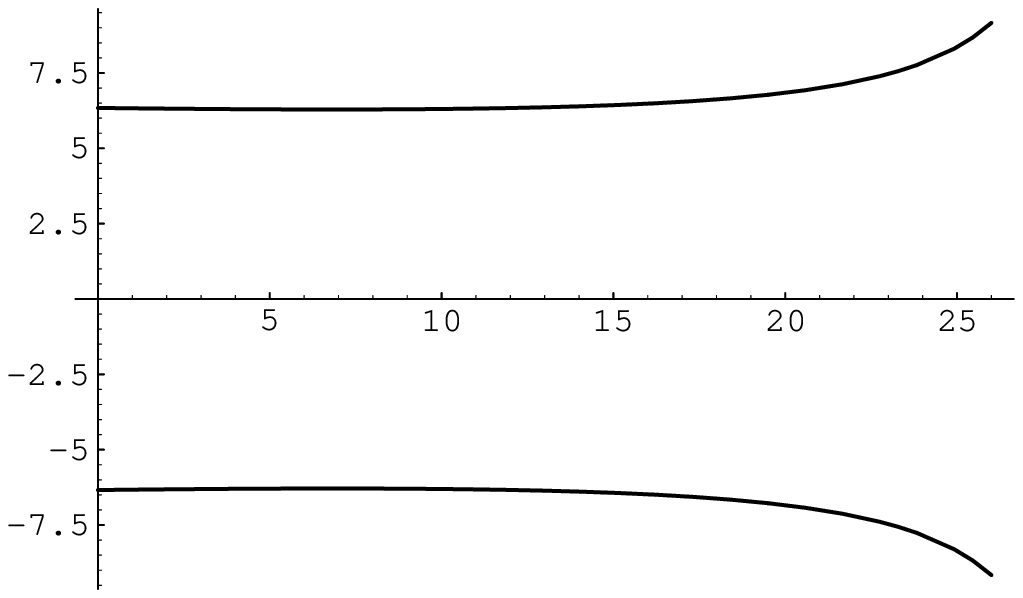}$$

\centerline{\sl Figure 1: Unitarity bounds on the $f^{0}(q^{2})$ form factor}
\vskip .5cm

However, we can impose some additional constraints.

\section{Constraints and Analysis}
\subsection{Constraints from Lattice data} 
To constrain our bounds, first we shall use a set of existing lattice 
results (for more technical details, please see \cite{ape}). The 
calculations of $f^{+}(t)$ and $f^{0}(t)$ are performed on the $APE$ 
machine, at $\beta = 6.0$, on a $18^{3}\times 64$ lattice using the Clover 
action. The extrapolation to the bottom mass is done using the scaling 
behavior predicted by the heavy quark effective theory which is fully 
justified as long as the calculations are performed in the vicinity of the 
zero-recoil point
($q^{2}\sim q_{max}^{2}$). The extrapolation to the light quark mass is 
quite smooth and unlikely to be a source of an important uncertainty within 
a 
present statistical accuracy. For the extrapolations to $q^{2} = 0$, several 
momenta were injected, but the question of this extrapolation left opened 
though (the results at small $q^2$ suffer from large errors). For our 
purpose, we take three results with the best precision.  
\vskip 3mm
\begin{center}
\begin{tabular}{|c||c|c|c|c|} \hline
$t \equiv q^{2}[GeV^{2}]$& 17.5 & 20.5 & 24.4 & 26.4 \\ \hline
$f^{0} $&  & .69$\pm$.10 & .64$\pm$.04 &.62$\pm$.04 \\ \hline
$f^{+}$ & 1.06$\pm$.65 & 1.5$\pm$.28 & 2.47$\pm$.26 &  \\ \hline
\end{tabular}
\end{center}
Now, let us define $g_{i}(z)=(1 - zz_{i}^{*})^{-1}$ ($i=1,2,3$ lattice 
points) and again construct $5 \times 5$ matrix whose determinant is 
positive. Generally, for the case $i=1,\ldots ,(n-2)$, this is the $n \times 
n$ matrix:
\vskip .03cm
\begin{equation}
\left| \begin{array}{ccccc}
\chi_{i} & f_{i}^{*}(z)\phi_{i}^{*} (z) & f_{i}^{*}(z_{1})\phi_{i}^{*} 
(z_{1})&  ... & f_{i}^{*}(z_{n})\phi_{i}^{*} (z_{n}) \\
 
f_{i}(z) \phi_{i} (z) & \frac{1}{1 - |z|^{2}} & \frac{1}{1 - zz_{1}^{*}}& 
... & \frac{1}{1 -  zz_{n}^{*}} \\
  &  &  &  &  \\
  & \cdots  &  \cdots & \cdots &  \\
f_{i}(z_{n})\phi_{i} (z_{n}) &\frac{1}{1 - z_{n}z^{*}}& \frac{1}{1 - 
z_{n}z_{1}^{*}} & ... &  \frac{1}{1 - |z_{n}|^{2}} \end{array} \right|  \geq 
0
\end{equation}
\vskip .3cm
This is the most compact form of this expression. As it can be noticed on 
Fig.2, the situation for $f^{0}(t)$ is very much improved: while the bare 
unitarity bounds give $|f^{0}(0)|\leq 6.3$, constraints imposed by the 
lattice results concentrated on the opposite end ($q^2 \simeq q^2_{max}$) 
give $-0.7 \leq f^{0}(0)\leq 
2.2$ . Errors are included (for the upper(lower) bounds we used the 
upper(lower) limits from the table).
$$ \epsfbox{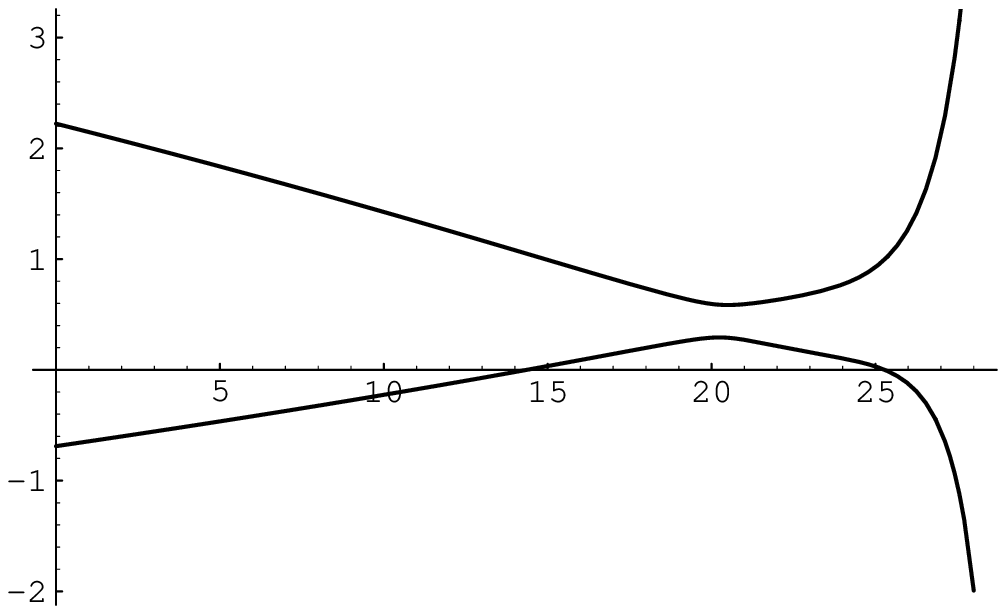}$$

\centerline{\sl Figure 2: Constrained unitarity bounds on the $f^{0}(q^{2})$ 
form factor;}\centerline{\sl 3 lattice values are taken (see text)}
\vskip .5cm

The situation for the more interesting $f^{+}(t)$ is not equally good: the 
lattice errors are larger, the Blaschke factor affects slightly  ( 
probably because they apply equally well for any value of residue). Still, 
the bounds are improved. Unfortunately, the analysis shows that using the 
constraint 
$f^{+}(0)=f^{0}(0)$ does not help for further strengthening our bounds. On 
Fig.3, we show these bounds against the simple pole behavior (dashed line, 
with $f^{+}(0)$ taken from Ref.\cite{wsb}). We 
see that $-0.7 \leq f^{+}(0)\leq 2.7$ which means that the `bare' allowed 
range is shrank by a factor of about $4$.
$$ \epsfbox{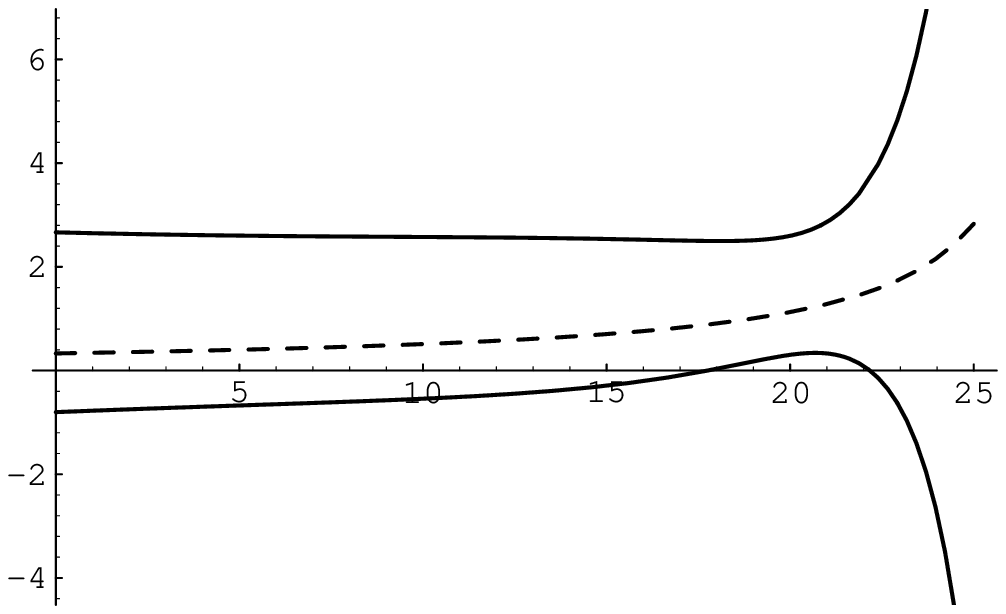}$$

\centerline{\sl Figure 3: Constrained unitarity bounds on the $f^{+}(q^{2})$ 
form factor;}\centerline{\sl 3 lattice values are taken (see text)}
\centerline{\sl Dashed line shows WBS model prediction}
\vskip .5cm
Now we can make an analysis of the methods and models applied so far in 
order to see if their predictions of $f^{+}(t)$ behavior fall within bounds 
in the whole region. For the moment, we want to get an information on the 
function $f^+(t)$. 

While LQCD computations give good results around zero-recoil point, the QSR 
are very successefull in the oposite region. To avoid the discussion on 
error estimations we will take some results of the form-factors that are not 
far from $q^{2} = 0$ point. To relate these two sets of results, we will 
make the fit, but with the essential physics of QCD incorporated.
Again we perform the conformal mapping:
\begin{eqnarray}
\frac{1+z}{1-z} = \sqrt{\frac{(m_{B}+m_{\pi})^{2}-t}{4Nm_{B}m_{\pi}}}
\end{eqnarray}
which is the same as the previous one for $N=1$. Here N is to be adjusted 
(see below).
The end points of the kinematic region are mapped as:
\begin{eqnarray}
t = 0  &\longmapsto&  z_{max}= \frac{m_{B}+m_{\pi}-2\sqrt{Nm_{B}m_{
\pi}}}{m_{B}+m_{\pi}+2\sqrt{Nm_{B}m_{\pi}}} \\
t = q^2_{max}  &\longmapsto&  z_{min}=-\left( \frac{\sqrt{N} - 1}{\sqrt{N} + 
1}
\right)
\end{eqnarray}
We concentrate on the $f^{+}(t)$. Proceeding the same analysis as above, we 
obtain:
\begin{equation}
\phi_{T}(z) = \frac{1}{4m_B\sqrt {3\pi Nm_{B}m_{\pi}}} \frac{(1 + z)^{2}}{(1 
- z)^{3}} \left( \frac{1}{\sqrt{N}} + \frac{1 + z}{1 - z}\right)^{3/2} \left(
\frac{m_{B}+m_{\pi}}{2\sqrt{Nm_{B}m_{\pi}}} + \frac{1 + z}{1 - z}
\right)^{-5}. 
\end{equation}

As long as $|z_{min}|$ and $|z_{max}|$ are smaller then one, $\bpi$ decay 
posseses a small kinematic expansion parameter. Since both $f^{+}(z) \to 
P_{*}(z)f^{+}(z)$ and 
$\phi_{T}(z) \to \phi_{T}(z)/\sqrt{\chi_{T}}$ are analytic on the unit disc, 
we can Taylor expand about $z = 0$:

\begin{equation}
f^{+}(z) = \frac{1}{P_{*}(z)\phi_{T}(z)} {\sum_{n=0}^{\infty}{a_{n}z^{n}}}
\qquad 
\end{equation}

After the above substitutions, the transformed inequality obtained from 
dispersion relations gives:

\begin{equation}
\frac{1}{2\pi i} \int_{\cal C} \frac{dz}{z} |\phi_{T}(z)f^{+}(z)|^{2} \leq 1
\end{equation}

Eqs.(25)and(26) give:
\begin{equation}
\sum_{n=0}^{\infty} |a_{n}|^{2} \leq 1
\end{equation}
This is a very important constraint which will be used in what follows.
Let us take first k-terms of this expansion: 
\begin{equation}
f_{k}^{+}(z) = \frac{1}{P_{*}(z)\phi_{T}(z)} \sum_{n=0}^{k}{a_{n}z^{n}}.
\end{equation}
k is to be chosen in such way that the truncation error is small. The 
expression for the maximum truncation error  can be obtained using the 
Schwarz inequality:

\begin{eqnarray}
\max{|f^{+}(z) - f_{k}^{+}(z)|} &\leq& \max {\frac{1}{P_{*}(z)\phi_{T}(z)} 
\sqrt{\sum_{n=k+1}^{\infty}{a_{n}^2}} \sqrt{\sum_{n=k+1}^{\infty}{z^{2n}}}} 
\\
&<& \max {\frac{1}{P_{*}(z)\phi_{T}(z)} \frac{|z^{k+1}|}{\sqrt{1 - z^{2}}}}
\end{eqnarray}
We could have applied this method to the previous case ($N=1$). However, in 
this case the physical range for $\bpi$ ( $0 \leq q^{2} \leq 26.4 GeV^{2}$) 
would have 
corresponded to $0 \leq z \leq 0.5183$. So, in order to have small 
truncation error, we should have had to include so many terms in the 
expansion 
that we could not evaluate such a large number of coefficients in expansion 
with just few points for the fit. 
However, there is a way to disentangle this problem by taking the optimal 
value of $N$. In our case, we obtain the optimal value $N = 3.45$, which 
gives $z_{min} = -0.300$ and $z_{max} = 0.258$. We repeat the 
analysis of the $f^{+}(t)$ constrained by the lattice results and our choice 
of $N$. To have the small truncation error ($\pm 0.03$) in the region of 
available lattice results, we have to take $k = 4$. Beside the three values 
from the table, we take the fourth one $f^{+}(19.75) = 1.19 \pm 0.24$ from 
the same simulation. The result of this analysis is very instructive. In the 
region 
around 23 $GeV^{2}$, we see that our bounds are severely constrained 
(Fig.4). 
$$ \epsfbox{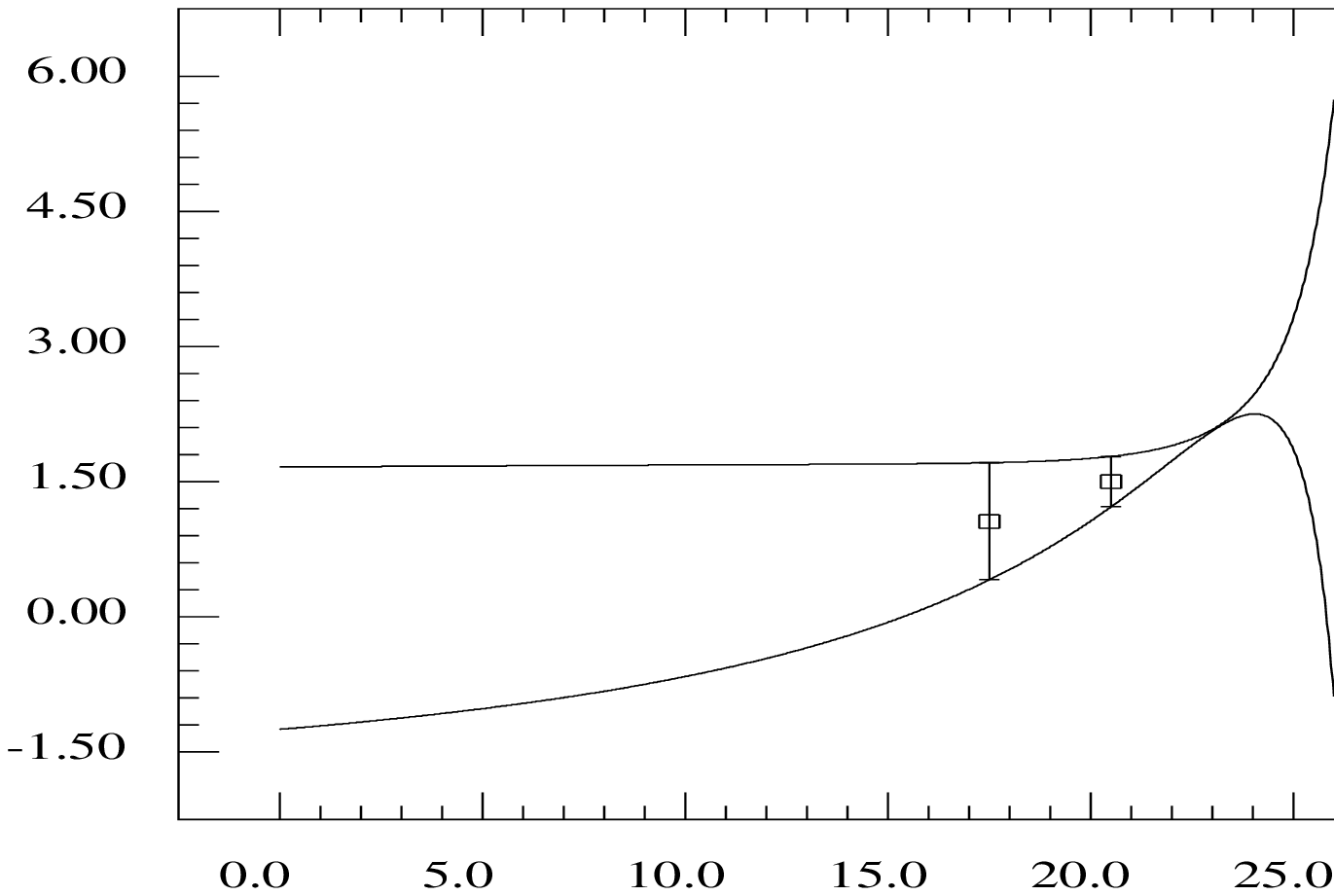}$$

\vspace{2.0cm}
\centerline{\sl Figure 4: More constrained bounds on the $f^{+}(q^{2})$ form 
factor;}\centerline{\sl 3 lattice values and N = 3.45 }
\vskip 0.5cm
It means that we can learn something about the value of the residue for this 
form factor. This point (23 $GeV^{2}$) is close to the zero-recoil point and 
is 
presumably dominated by the lightest state which couples to the weak 
current $V_{\mu}$, namely the $B^{*}$ ($M_{B^*}^2 = 28.3 GeV^2$). 
 
\begin{equation} 
f^{+}(q^{2}) = \frac{F_{*}}{m_{B^{*}}^{2} - q^{2}}
\end{equation}

($F_{*}$ stands for the residue), we obtain $F_{*} = (10.84\pm.03) GeV^{2}$. 
Of course from dispersion relation for this form factor, we have also the 
contribution of continuum, that we take to be negligible relative to the 
pole one. If pole-dominance is assumed, this value of residue gives 
$f^{+}(0) = 0.38 \pm .03$.

\subsection{Inclusion of QSR data} 

However, our kinematic region is large and we can not that easily conclude 
about the value $f^{+}(0)$,nor the functional dependence of $f^{+}(t)$. To 
get some additional information, we shall take advantage of the knowledge of 
well estimated values of this 
form factor in the region of the small $q^2$. Specifically, we take the 
updated results of 
the light-cone sum rules which proved to be very stable in the range $0 
\leq q^{2} \leq 15\ GeV^{2}$. In what follows, we take only central values 
in both sets of results ( QSR and LQCD). Again, the goal is to learn more 
about $f^{+}(t)$ behavior.
For the fit, we use eight points which ensure that the truncation error is 
less than 0.03. Here, we recall the important constraint (27) which 
eventually reduce the number of coefficients contributing to (28) to four. 
Completed by $QSR$ data, the set for the fit is:

\begin{center}
\begin{tabular}{|c||c|c|c|c|c|c|c|c|} \hline
$t \equiv q^{2}[GeV^{2}]$& 0 & 4 & 8 & 10 & 15 & 17.5 & 20.5 & 24.4 \\ \hline
$z$&0.2584 & 0.2239& 0.1829 &0.1590&0.0856 & 0.0380& -0.0348&-0.1775\\ \hline
$f^{+}$ & 0.29 & 0.35& 0.45& 0.52& 0.83&1.06 & 1.5 & 2.47  \\ \hline
\end{tabular}
\end{center}

The coefficients $a_{n}$ are:

\begin{center}
\begin{tabular}{|c||c|c|c|c|c|c|c|c|} \hline
$a_{0}$&$a_{1}$  & $a_{2}$ & $a_{3}$ & $a_{4}$ \\ \hline
3.33$\cdot 10^{-3}$&0.36$\cdot 10^{-3} $& -33.29$\cdot 10^{-3}$& -5.32$\cdot 
10^{-3}$ &233.11$\cdot  10^{-3}$\\ \hline
\end{tabular}
\end{center}

On Fig.5a, we plot $f^{+}_{1}(q^2)$ (with constraint (27)) versus 
$f^{+}_{2}(q^2)$ (without it). On Fig.5b $f^{+}_{1}(q^2)$ versus pole 
behavior, and on Fig.5c $f^{+}_{2}(q^2)$ versus double pole behavior is 
ploted. For the pole we take (31) with the residue given below the equation, 
and for double pole we take. parameters from Ref.\cite{uk} )

\vspace{-5mm}
$$ \epsfbox{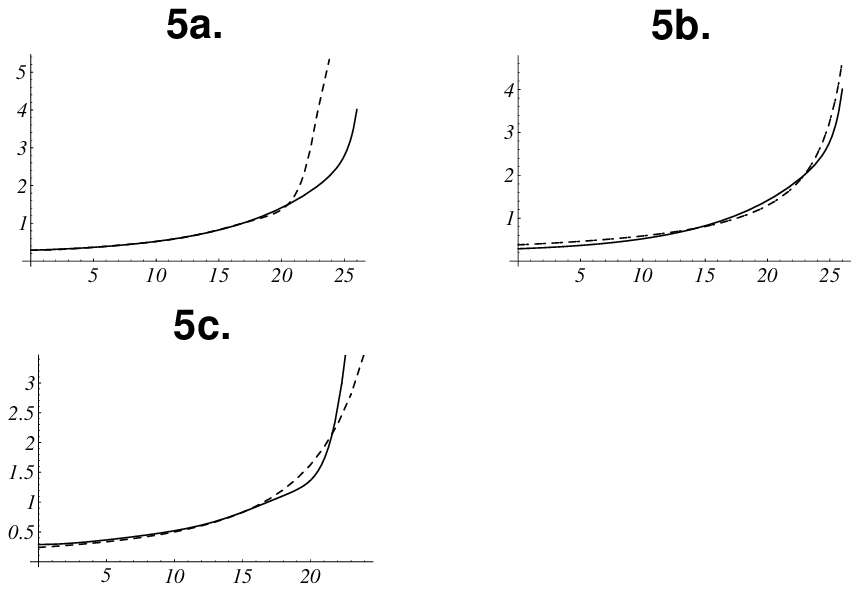}$$
\parbox[b]{16cm}{\vspace{-1cm}
\begin{center}\sl Figure 5: a) $f^{+}_{1}(q^{2})$ (fit with data from the 
table and constraint $\sum_{n=0}^{8} |a_{n}|^{2} \leq 1$).
 Dashed curve $f^{+}_{2}(q^{2})$ is corresponding unconstrained fit.
\sl b) $f^{+}_{1}(q^{2})$ versus $f^{+}_{pole}(q^{2})$ (dot-dashed)
c) $f^{+}_{2}(q^{2})$ versus 'double-pole' behavior  (dashed)
\end{center}}

\vskip 0.5cm
Let us briefly comment the figures. 
On the first plot, we see the effect of the constraint 
(27). It suppresses the form factor at large $q^{2}$ and constrains it 
to follow rather pole dominance. This is evident on the second plot. There 
is no compelling theoretical justification for the usage of the 
pole-dominance ansatz 
(except for the points close to the zero-recoil). However, from this 
analysis, we see that this behavior is strongly favored. Still, there is no 
convincing arguments to neglect completely the contributions of the 
singularities above the threshold. According to the present data, we can not 
estimate deviations from the simple pole behavior. Lattice data which are 
concentrated around $q^{2}_{max}$ should be much more precise for this 
analysis. There is a permanent tendency in reducing all posible sources of 
errors (statistic and systematic) and we hope that next simulations will 
provide us the data for a pertinent analysis of the excited states above the 
threshold. As we already mentioned, for the extrapolation of the lattice 
data some ans\"{a}tze are usually considered. These hypothesis must satisfy 
the kinematical constraint ($f^{0}(0)=f^{+}(0)$), as well as the scaling 
relations in the heavy quark limit, $M\to \infty$ ($f^{0}(q^{2}_{max} \sim 
M^{-1/2}$,$f^{+}(q^{2}_{max} \sim M^{1/2}$). Then, the popular ansatz is:  
\begin{equation}
f^{i}(q^{2}) = \frac{f^{i}(0)}{\left( 1 - \frac{q^{2}}{M_{i}^{2}}
\right)^{n_{i}}}
\end{equation}
where $n_{+} = n_{0} + 1$. The $UKQCD$-results suggested double pole 
behavior. We take their parameters and plot $f^{+}(q^{2})$ \cite{uk} versus 
`unconstrained' fit. According to our analysis, the double pole behavior is 
not 
acceptable since the constraint (27) bends the curve and favours rather the 
simple-pole one.

\section {Summary}

There is no nonperturbative calculational method to evaluate form-factors in 
the whole kinematic range for the semileptonic heavy to light decays. Even 
more, there is no fully reliable principles which could help. Hence, all 
that can be done is to use a phenomenological ansatz. In this paper we 
wanted to attack the problem the other way around. The calculations 
performed in 
the unphysical region are related to the form factors of interest via 
crossing symmetry and dispersion relations. From the derived set of 
inequalities, using the conformal mapping, we obtained unitarity bounds on 
the form factors. We have shown that such bounds are not restrictive. To 
narrow 
the range of allowed values of the form factors, we used results from the 
simulation on the lattice ($APE$). Additional constraint in this method 
comes from the optimal choice of parameter $N$. As a result we get a very 
narrow strip of allowed values of $f^{+}(q^{2})$ around 23.2 $GeV^{2}$. To 
answer the question about the $q^{2}$ behavior of the form-factor, we used 
the light cone QSR results. Having the essential physics incorporated, we 
perform a simple fit limiting ourselves to the central values predicted by 
the
two methods. As a result, we obtained the curve which favors behavior 
dominated by the $B^{*}$. From $f^{+}(23.2\  GeV^{2}\,)$ we obtain the value 
of 
residue\footnote{this value of the residue is in a good agreement with the 
value of $g_{B^*B\pi}$ obtained in Ref. \cite{bra}}  $F_{*} = (10.84\pm.03) 
GeV^{2}$ which corresponds, if the pole dominance is valid in the whole 
kinematic range,to  
$f^{+}(0) = 0.38 \pm .03$. Deviations from this law near $q^{2}_{max}$ are 
present and they come from the excited $B^{*}$ states which are located 
above 
the threshold. For the proper estimation of their contributions more precise 
lattice data are needed. The other way could be to use the $D \rightarrow \pi
\ell \bar\nu_{\ell}$  form-factors and to extrapolate to $\bpi$ armed by 
$HQS$. Then the perturbative contributions must be calculated at $Q^{2} = 16 
GeV^{2}$ and $\cal O(\alpha_{s})$ corrections are large.  
In our analysis we used the only two nonperturbative methods so that the 
bounds and outlined behavior in $q^{2}$ are model independent.

\vskip 1.5cm
{\bf Acknowledgement}
\vskip 1.0cm

It is a pleasure to thank A. Le Yaouanc and J. P. Leroy for useful and 
motivating discussions and also M. Crisafulli and A. Khodjamirian for the 
form-factors data.

\end{document}